\documentclass[aps,amsmath,amssymb,twocolumn,preprintnumbers,superscriptaddress,showpacs]{revtex4}

\usepackage{color}
\usepackage{graphicx}
\usepackage{dcolumn}
\usepackage{bm}
\usepackage[latin1]{inputenc}
\usepackage{epsfig}                    

\begin{document}

\title{Magnetic ordering in electronically phase separated Sr/O co-doped La$_{2-x}$Sr$_x$CuO$_{4+y}$}

\author{L.~Udby}
\affiliation{Materials Research Division, Ris\o{} DTU, Technical University of Denmark, DK-4000 Roskilde, Denmark}

\author{N.H.~Andersen}
\affiliation{Materials Research Division, Ris\o{} DTU, Technical University of Denmark, DK-4000 Roskilde, Denmark}
\email{nhes@risoe.dtu.dk}

\author{F.C.~Chou}
\affiliation{Center for Condensed Matter Sciences, National Taiwan University, Taipei 10617, Taiwan}

\author{N.B.~Christensen}
\affiliation{Materials Research Division, Ris\o{} DTU, Technical
University of Denmark, DK-4000 Roskilde, Denmark}
\affiliation{Laboratory for Neutron Scattering, ETHZ \& PSI,
Ch-5232 Villigen PSI, Switzerland}
 \affiliation{ Nanoscience
Center, Niels Bohr Institute, University of Copenhagen, DK-2100
Copenhagen, Denmark}

\author{S.B.~Emery}
\affiliation{Department of Physics, University of Connecticut U-3046,2152 Hillside Road, Storrs, Connecticut 06269-3046, USA}

\author{K.~Lefmann}
\affiliation{Materials Research Division, Ris\o{} DTU, Technical University of Denmark, DK-4000 Roskilde, Denmark}
\affiliation{Nanoscience Center, Niels Bohr Institute, University of Copenhagen, DK-2100 Copenhagen, Denmark}
\affiliation{ESS-Scandinavia, University of Lund, Stora Algatan 4, 22350 Lund, Sweden}

\author{J.W.~Lynn}
\affiliation{NIST Center for Neutron Research, Gaithersburg, MD 20899, USA}

\author{H.E.~Mohottala}
\affiliation{Department of Physics, University of Connecticut U-3046,2152 Hillside Road, Storrs, Connecticut 06269-3046, USA}

\author{Ch.~Niedermayer}
\affiliation{Laboratory for Neutron Scattering, ETHZ \& PSI, Ch-5232 Villigen PSI, Switzerland}

\author{B.O.~Wells}
\affiliation{Department of Physics, University of Connecticut U-3046,2152 Hillside Road, Storrs, Connecticut 06269-3046, USA}

\date{\today}

\begin{abstract}
We present results of magnetic neutron  diffraction experiments on the co-doped super-oxygenated La$_{2-x}$Sr$_x$CuO$_{4+y}$ (LSCO$_{\text{+O}}$) system with $x=0.09$. The spin-density wave has been studied and we find long-range incommensurate antiferromagnetic order below $T_N$ coinciding with the superconducting ordering temperature $T_c=40$ K. The incommensurability value is consistent with a hole-doping of $n_h\approx \frac{1}{8}$, but in contrast to non-superoxygenated La$_{2-x}$Sr$_x$CuO$_{4}$ with hole-doping close to $n_h\approx \frac{1}{8}$ the magnetic order parameter is not field-dependent. We attribute this to the magnetic order being fully developed in LSCO$_{\text{+O}}$ as in the other striped lanthanum-cuprate systems.\\

\end{abstract}

\maketitle

\section{\label{sec:intro} Introduction}
The presence of an inhomogeneous charge concentration in cuprate superconductors has become increasingly obvious in recent years. The most dramatic experiments showing local density of state variations have been performed using Scanning Tunnelling Microscopy/Spectroscopy \cite{lang_N415, howald_PRB64, kohsaka_S315}. There has also been an increasing number of other experiments that are best explained by invoking an inhomogeneous electronic structure \cite{zimmermann_EPL41,dordevic_PRL91, singer_PRL88, wakimoto_PRL98}. For most of these experiments the charge variations appear to be short ranged, associated with a length scale of only a few nanometers at most. However, for the special cases of oxygen doped La$_2$CuO$_{4+y}$ (LCO$_{+\text{O}}$) or oxygen co-doped La$_{2-x}$Sr$_x$CuO$_{4+y}$ (LSCO$_{+\text{O}}$), muon and superconducting quantum interference techniques suggest that the electronic inhomogeneity moves beyond such local variations to form fully phase separated regions \cite{savici_PRB66, mohottala_NM5}. For both cases, with hole concentrations ($n_h$) between 0.125 and 0.16 per Cu site, samples at low temperatures spontaneously form separate regions of a magnetic phase consistent with $n_h$=1/8 that is not superconducting and an optimally doped superconductor with $n_h$=0.16 that is not magnetically ordered. The driving force for this phase separation appears to be interactions between the doped holes themselves rather than any specific O or Sr chemistry\cite{mohottala_NM5}.

The full implications of this complete phase separation are still to be determined both theoretically and empirically. One area where phase separation should certainly be important is for effects associated with competing order parameters. A pronounced case of such an effect is the large magnetic field enhancement of the ordered moment in underdoped La$_{2-x}$Sr$_x$CuO$_{4}$ (LSCO) superconductors as measured by neutron diffraction. A series of experiments have shown that samples with x$<$1/8 have an incommensurate (IC) antiferromagnetic (AFM) elastic diffraction peak that grows substantially with the application of a magnetic field \cite{chang_PRB78, lake_N415}. Samples with x$\gtrsim$0.14 have no elastic magnetic peak in zero applied field, but such a peak appears at a critical field, $H_c$, and then grows in intensity as the field increases above that  \cite{khaykovich_PRB71, chang_PRB78}. For samples doped very close to $x=1/8$, and for which suppression of the superconducting $T_c$ is also observed, a strong magnetic peak exists at zero field with less enhancement from the application of a field. Samples of La$_{1.88}$Sr$_{0.12}$CuO$_4$ still show a small field enhancement \cite{katano_PRB62}, while samples of La$_{1.48}$Nd$_{0.4}$Sr$_{0.12}$CuO$_4$ (LNSCO) and La$_{7/8}$Ba$_{1/8}$CuO$_{4}$ (LBCO) have a fully developed magnetic moment and no - or very small - field enhancement\cite{chang_PRB78,fujita_JPCS51,wen_PRB78}. We will hereafter refer to samples with fully developed magnetic order and no field enhancement within 13.5 T applied field as true 1/8 samples.\\
\indent A widely used theory for the intensity enhancement by application of an external field has been developed by Demler, Sachdev, and Zhang \cite{demler_PRL87}. This theory (DSZ) describes the cuprates as having coexisting but competing magnetic and superconducting order parameters. The functional form for the magnetic peak intensity versus field appears to fit existing data well and the predicted phase diagram appears to qualitatively match measurements for samples which are not true 1/8. However, the observation that true 1/8 samples have no field enhancement does not match the predictions of the DSZ paper. Of natural interest is how this theory of competing but coexisting order parameters might be adapted for related samples which appear to have fully separated order parameters that do not coexist at the same location in the sample.\\

In the present work we study the details of the elastic magnetic scattering of a LSCO$_{\text{+O}}$ single crystal by neutron diffraction. The phase separation, phase fractions, zero field ordered moment, and flux pinning have been carefully measured previously using muon spin resonance ($\mu$SR) and bulk susceptibility measurements \cite{mohottala_NM5, mohottala_PRB78}. Here we establish that the magnetic phase is IC AFM and long-range ordered with the same incommensurability as the true 1/8 samples of LNSCO and LBCO. This is surprising since the superconducting transition temperature is not suppressed in our sample but instead is very high ($T_c=40$ K) and coinciding with the ordering temperature of the IC AFM. Field and temperature dependence of the IC AFM peak intensity is also presented.
We discuss the field dependence in relation to the DSZ model, pointing out that some development is necessary to account for samples where the magnetic and superconducting phases fully separate rather than coexist.

\section{\label{sec:method} Methods}
Our sample is a co-doped single-crystal with Sr content $x=0.09$ (LSCO$_{\text{+O}}^{x=0.09}$). It has mass $m=0.429$ g. It was grown by the travelling solvent floating zone method in a mirror furnace. Additional oxygen was introduced using wet electrochemical methods as presented previously\cite{wells_ZP100}.
Previous studies of this particular crystal showed onset for both superconductivity and magnetism at 40 K\cite{mohottala_NM5}. Only one superconducting and one magnetic phase were detected, and each of these two phases occupy close to 50\% of the volume as measured by $\mu$SR.

The neutron diffraction measurements were performed at the cold neutron triple-axis spectrometers RITA-II at PSI and SPINS at NIST. In  both experiments we used 5 meV neutrons, 40' collimation before the sample and Be filter before the analyzer. Error bars in this manuscript are statistical in nature and represent one standard deviation.

RITA-II has the special feature of a seven blade PG$(002)$ analyzer making it possible to monitor seven different reciprocal space points at the same time and energy-transfer, the so-called monochromatic imaging mode \cite{lefmann_PhysicaB283,bahl_NIMP226,bahl_NIMP246}. This enables simultaneous measurements of peak and background, which have proven very useful since the weak magnetic signal requires very long counting times of the order of hours. The size of the sample and the width of the analyzer blades result in an effective horizontal collimation of 40' between the sample and each analyzer blade.

The LSCO system is subject to twinning when in the low temperature orthorhombic (LTO) state. In terms of the $F4/mmm$ unit cell for the high temperature tetragonal (HTT) structure the twinning is along (110) and ($1\bar{1}0$). The orthorhombic axes are almost parallel to the  $F4/mmm$ axes and the twinning gives up to four peaks in the LTO phase for each peak in the high temperature tetragonal (HTT) phase\cite{braden_PC}. In our crystal even at low temperatures in the  LTO state\cite{udby_structure}, the difference between lattice constants $a=5.318(4)$\AA{} and $b=5.337(6)$\AA{} gives rise to only a tiny transversal splitting of $\alpha=0.10(4)^\circ$ across the $H$ and $K$ axes.

All references to crystallography are in the LTO $Bmab$ notation unless explicitly stated otherwise.

\section{\label{sec:results} Results }
\begin{figure}
\includegraphics[width=\columnwidth]{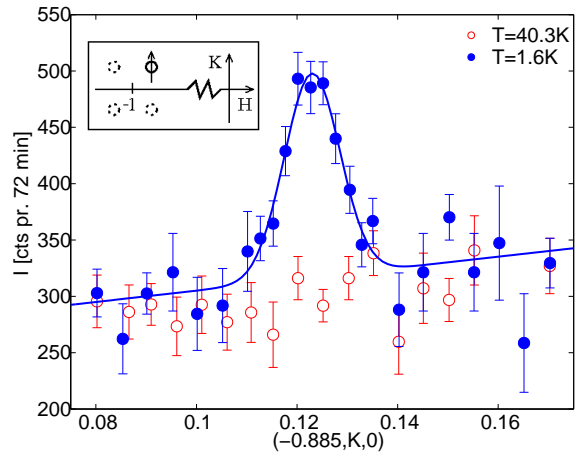}
\caption{\label{fig:ICpeak}(Color online) Scans in reciprocal space through \mbox{(-0.885,0.123,0)}  along the $K$-direction. Scans are taken above and below $T_N$ in zero applied field, the solid line is a fit to the data as explained in section \ref{sec:discuss}. The inset shows the peak position and scan direction.}
\end{figure}

Previous $\mu$SR studies\cite{mohottala_NM5} of our LSCO$_{\text{+O}}^{x=0.09}$ crystal have shown a strongly damped oscillatory behavior with $\nu=3.33(8)$MHz. This corresponds to an internal field of 24.7 mT which is about 2/3 of the value observed in the undoped compound La$_2$CuO$_4$ (LCO)\cite{budnick_EPL5}.

Our neutron diffraction studies find IC AFM elastic peaks at the same scattering vectors as in superoxygenated LCO$_{\text{+O}}$ crystals\cite{lee_PRB60,lee_PRB69}. Scans through the incommensurate point are shown in Fig. \ref{fig:ICpeak} at base temperature and just above the magnetic transition temperature $T_N$.
The peak incommensurability is $\delta_H=0.1150(18)$ r.l.u. and $\delta_K=0.1230(5)$ r.l.u., respectively, which gives a distance $\delta=0.198(3)$\AA$^{-1}$ from the AFM point. This corresponds to an incommensurability of $\delta_T=0.119(2)$ r.l.u. in pseudo-tetragonal notation \footnote{$a_T=b_T=\frac{a+b}{2\sqrt{2}}=\frac{5.328(4)}{\sqrt{2}}=3.767(3)$\AA{}}, which is consistent with a hole-doping of $n_h\approx 1/8$ according to the Yamada-plot \cite{yamada_PRB57}  and consistent with the $\delta\approx0.12$ of LBCO\cite{fujita_PRB70} and LNSCO\cite{chang_PRB78}. The incommensurate peaks are similar to or sharper than the previously reported instrumentally resolved ones in LCO$_{\text{+O}}$\cite{lee_PRB60,lee_PRB69,khaykovich_PRB66,khaykovich_PRB67}.

It has been checked by multiple tests that the peak width, position and amplitude do not depend on cooling rate or cycle.

\begin{figure}
\includegraphics[width=\columnwidth]{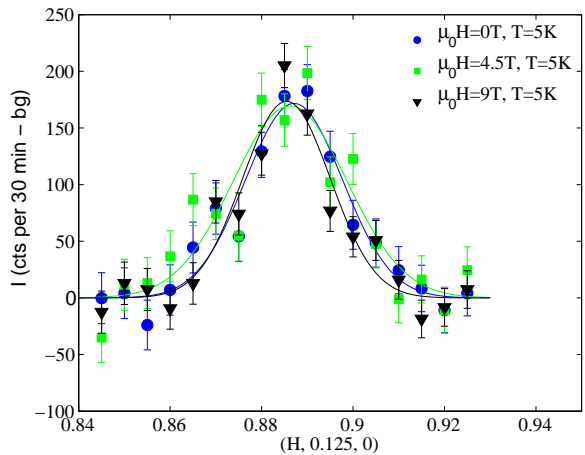}
\caption{\label{fig:ICpeakfielddep}(Color online) H-scans through the IC peak position at various fields. The experimental data have been fit to a Gaussian on sloping background. In the figure the data points and their Gaussian fits (solid lines) are shown after subtracting their sloping background (940 cts/30 min at the peak position). All fitted Gaussian parameters are the same within error bars for the different fields.}
\end{figure}

\begin{figure}
\includegraphics[width=\columnwidth]{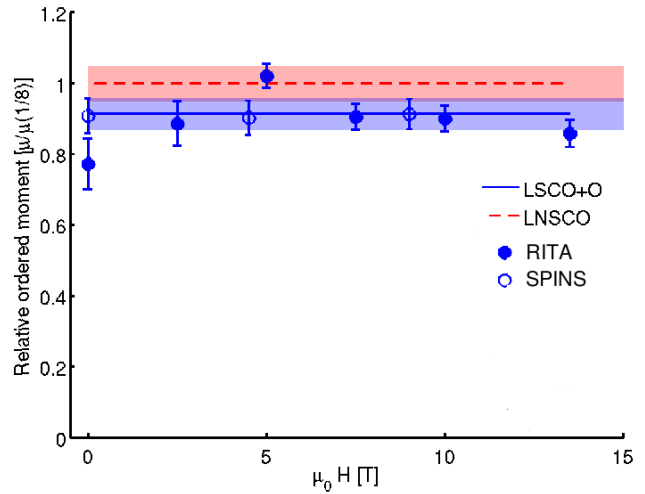}
\caption{\label{fig:ICfielddep}(Color online) Internal magnetic moment (square root of SDW peak intensities) as a function of applied field. Closed symbols are data from RITA-II (measuring one point at the peak position) and open  symbols are from SPINS (full momentum scan fitted to a Gaussian) scaled to weighted average of RITA-II data. A constant fit to the data (solid line) is shown relative to the internal moment of LNSCO from \cite{chang_PRB78} (dashed line). Shaded areas indicate the error related to the determination of the local ordered moment from $\mu$SR.}
\end{figure}

As is shown in Fig. \ref{fig:ICpeakfielddep}, application of a field does not shift or broaden the IC peak within the error bars, making it possible to monitor the intensity as a function of applied field in few-point scans.
In imaging mode at RITA-II the central blade is used for measuring the peak amplitude whereas the other blades measure the background which is found to be field independent. Data for applied fields up to 13.5 T are shown in Fig. \ref{fig:ICfielddep}. Simply fitting the peak intensity to a constant describes our data quite accurately meaning that we observe no field effect within this field range. This is similar to the anomalous behavior of the other true 1/8 samples such as LNSCO and LBCO. In addition, the LSCO$_{\text{+O}}$ system also has the same $\mu$SR response as LBCO and LNSCO, which is a strongly damped oscillatory behavior with a frequency $\nu\approx3.5$ MHz corresponding to a local ordered moment of $\sim0.35\mu_B$\cite{mohottala_NM5,nachumi_PRB58,chang_PRB78}. The neutron scattering data are proportional to the ordered spin moment squared. Therefore, in order to compare with $\mu$SR results, our data have been presented after taking the square root of the background subtracted measured intensities and scaling to LNSCO muon data in \cite{chang_PRB78}.

\begin{figure}
\includegraphics[width=\columnwidth]{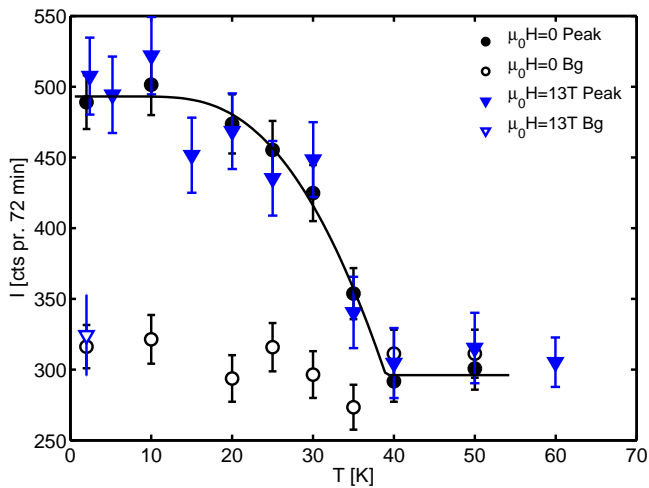}
\caption{\label{fig:ICtemp}(Color online) Temperature dependence of the IC SDW peak intensity at $\mu_0 H=$0 (in cryostat) and $\mu_0 H=$13.5 T (in magnet, after field-cooling) respectively. The $\mu_0 H=$13.5 T data are scaled (from the ratio of Bragg peak intensity in the cryostat and magnet respectively) and subtracted a constant background. The solid curve is a guide to the eye.}
\end{figure}

The temperature dependence of the IC spin density wave (SDW) peak intensity is shown in Fig.  \ref{fig:ICtemp} for both 0 T and 13.5 T applied field. From a linear mean-field fit we find a magnetic ordering temperature of $T_N=40(4)$ K for both the 0 T and 13.5 T data. The magnetic transition temperature is well within the experimental uncertainties of the measured bulk superconducting transition temperature.

\section{\label{sec:discuss}Discussion}
We first consider the IC AFM peak of Fig. \ref{fig:ICpeak}. We find that the peak has $FWHM_{IC}=0.0130(14)$ r.l.u. by a simple Gaussian fit on sloping background, which is slightly broader than the resolution at the $(-1,0,0)$ position having $FWHM_{res}=0.0101(1)$ r.l.u.\footnote{This was measured by 2. order scattering off the structural $(-2,0,0)$ peak (without Be filter) but the resolution is similar in angular space}.
However, even if we take the small excess width with respect to the resolution width to be due to finite-size broadening, the SDW correlation length will still exceed 400\AA{}\footnote{If the peak is fitted to a Gaussian the combined width is the sum of the internal width $\sigma$ and the resolution in quadrature giving $\sigma=\sqrt{(0.0130*\frac{2\pi}{5.318})^2-(0.0101*\frac{2\pi }{5.337})^2 }=0.00972$ \AA$^{-1}$. The estimate of the correlation length can be either $2\pi /\sigma=646$\AA{} or by the Scherrer formula $\frac{K\lambda}{B\cos{\theta}}=\frac{0.9\cdot4.045}{0.00972\cdot 0.9404}=414$\AA{}  with $B=2\tan^{-1}\frac{\sigma}{2q_{IC}}$. The Scherrer dimensionless constant $K$ is dependent on domain-geometry but the value of 0.9 used here is correct within 10\% error. The error on the Lorentzian width on fitting to a Voigt is as large as the value itself, hence it has not been considered fitting to a Voigt in the main text.}.
Hence it is reasonable to conclude that the IC peak is close to resolution limited and expressing long range SDW order.
 The unconvoluted width of our IC AFM peak is the same as the width of the IC AFM peak of LSCO $x=0.12$ and LNSCO \cite{chang_PRB78} which are resolution limited. Thus the correlation length is maximum at $x\sim1/8$ whereas it decreases in LSCO when doping departs from $x=1/8$\cite{chang_PRB78,fujita_PRB65}.

Imposing a field does not change the intensity, correlation length, incommensurability or transition temperature of the magnetic phase and it is hence reasonable to conclude that the magnetic state in the part of the sample with $n_h=1/8$ is already fully developed in zero field.

This is in contrast to LSCO $x=0.12$ for which neutron diffraction studies have shown that the field enhancement matches the functional form of the DSZ theory. In this case SC and the SDW coexist but the local ordered magnetic moment is not fully developed in zero applied field. Imposing a field, however, pushes it towards the local ordered moment of the true 1/8 state\cite{chang_PRB78}.

Enhancement up to a factor of two of the elastic IC AFM peak at moderate fields ($<$ 8 T) has also been observed by neutron diffraction in non-Sr doped LCO$_{\text{+O}}$ crystals when they were cooled slowly enough for the excess oxygen to order\cite{lee_PRB69}. The oxygen ordering is observable by the concomitant staging superstructure.
At the time of writing the authors of \cite{lee_PRB69} did not consider their crystal to be macroscopically phase separated. Given more recent developments, and the fact that the reported magnetic and superconducting properties of that crystal are very similar to ours, it seems likely that the sample used in that report was indeed phase separated in a manner similar to the sample we present here. Hence in the following discussion we will assume that this is the case. It is however important to bear in mind that the value of the magnetic volume fraction of otherwise similar LCO$_{\text{+O}}$ crystals can vary alot 
\footnote{As an example the LCO$_{\text{+O}}$ crystals of \cite{mohottala_NM5} and \cite{emery} were prepared by a similar electrochemical method and both have $T_c=T_M=40$ K but the first had magnetic volume fraction of 66(5)\% whereas the second only had 30(5)\% measured in zero applied field and by the same $\mu$SR method after similar cooling conditions. We should mention that the first crystal is very small (m=0.025 g) and hence no attempt has been made so far to measure the field dependence of the magnetic signal.}.
 
One possible explanation for the increase in the magnetic signal in \cite{lee_PRB69} could be that the local magnetic moment in this sample was not saturated in zero field and after slow cooling. However, from our previous $\mu$SR work we know that highly oxygenated LCO$_{\text{+O}}$ crystals have a fully developed local magnetic moment at zero applied field\cite{mohottala_NM5}, so the field effect of LCO$_{\text{+O}}$ is probably not explained by an unsaturated magnetic moment.
We speculate instead that it is due to the ability of LCO$_{\text{+O}}$ to convert SC regions into SDW in the case where the excess oxygen is ordered. The mechanism behind this might be similar to what is found in YBa$_2$Cu$_3$O$_{6+x}$ (YBCO). In YBCO \cite{zimmermann_PRB68,uimin_PRB50} oxygen ordering facilitates itinerant doped holes thereby favoring SC, whereas oxygen disorder does not favor SC. In this scenario SC would be favored in slowly cooled oxygen ordered LCO$_{\text{+O}}$ at least at zero applied field. Applying moderate fields hereafter allows SDW regions to grow to a plateau volume. In LCO$_{\text{+O}}$ oxygen disorder can be introduced by fast cooling. In LSCO$_{\text{+O}}$, which can be viewed as doping LCO$_{\text{+O}}$ with Sr, the homogeneously distributed Sr anti-correlates to the excess oxygen, creating an increasingly disordered oxygen distribution with increasing Sr content \cite{udby_structure}. Thus following this scenario, in LSCO$_{\text{+O}}^{x=0.09}$ and fast-cooled LCO$_{\text{+O}}$, SC regions are not particularly favored over SDW regions even at zero applied field. This can explain why we see little or no enhancement of the magnetic signal by application of a field in the LSCO$_{\text{+O}}$ system.

Let us now consider the volume of our sample with $n_h=0.16$ separately and treat it within the DSZ frame. Then keeping our high $T_c$ of 40 K in mind, probably the critical field needed to actually enhance the magnetic signal in LSCO$_{\text{+O}}$ would be at least as large as that of LSCO $x=0.16$ (optimally doped). According to the DSZ phase diagramme, the critical field increases rapidly with increasing $x$. The fact that the critical field of $x=0.145$ is already 7 T \cite{chang_PRB78}, suggests that the critical field for $x=0.16$ would not be within our experimental reach.

The total outcome considering both phases would be that neither of them would show significant field enhancement of the magnetic signal in moderate fields, which is indeed what we observe.

There was no evidence in our co-doped LSCO$_{\text{+O}}$ crystal of any superconducting phase with $T_c$ different from 40 K\cite{mohottala_NM5}, nor did we observe any signs of the Néel antiferromagnetic order observed in LCO\cite{budnick_EPL5} as well as in the hole-poor phase of non-Sr-doped LCO$_{\text{+O}}$\cite{ansaldo_PRB40}. This is corroborating evidence for the suggestion\cite{mohottala_NM5} that in the region of the LSCO+O phase diagram to which our sample belongs, there exist only two stable ground states:  The optimally doped superconducting phase and the "true 1/8" magnetically ordered SDW phase.

\section{Conclusion}
We conclude that the magnetic phase in our LSCO$_{\text{+O}}$ crystal consists of fully developed long-range SDW order corresponding to the SDW of the 1/8 compounds LBCO and LNSCO with the same incommensurability $\delta\approx0.12$, local Cu$^{2+}$ moment $\sim 0.35\mu_B$ and lack of field effect. The regions occupied by the SDW are large, at least 400 \AA{}, meaning that below 40 K large SDW islands (or patches correlated over large distances) form simultaneously with the optimally doped SC in the rest of the sample.

Since there is no enhancement of the IC AFM peak with application of external field, we conclude that the LSCO$_{\text{+O}}$ system has a fully developed magnetic phase which cannot be expanded at least below 13.5 T applied field. In the slow-cooled highly oxygen-doped LCO$_{\text{+O}}$ system with the same magnetic structure there is clearly a (large) enhancement of the SDW peak with field. We speculate that the discrepancy with respect to our system is due to the ability of the (slow-cooled) LCO$_{\text{+O}}$ system to convert SC volume into SDW volume by application of an external field.  This ability might be related to the degree of oxygen ordering in the sample.

Our neutron scattering measurements support that co-doping facilitates long-range electronic phase separation below  $T_N=T_c =40K$ in two phases, 40 K SC and true 1/8 SDW, whose relative amounts are only determined by the total hole content.

\begin{acknowledgments}
We would like to thank Brian M. Andersen for many helpful
discussions. This work was supported by the Danish Agency for
Science Technology and Innovation under the Framework Programme on
Superconductivity and the Danish Research Council FNU through the
instrument center DANSCATT. Work at the University of Connecticut
was supported by the U.S. Department of Energy under Contract No.
DE-FG02-00ER45801. Part of this work was performed at the Swiss
Spallation Neutron Source, Paul Scherrer Institute, Villigen,
Switzerland. Work on SPINS is supported in part by the National
Science Foundation under Agreement No. DMR-0454672.
\end{acknowledgments}


\end{document}